# Unveiling the role of dielectric trap states on capacitively coupled radio-frequency plasma discharge: dynamic charging behaviors

Shu Zhang[1], Guang-Yu Sun[2], Arnas Volčokas,[2] Guan-Jun Zhang[1*] and An-Bang Sun[2*]

[1]*State Key Laboratory of Electrical Insulation and Power Equipment, School of Electrical Engineering, Xi'an Jiaotong University, Xi'an, Shaanxi, 710049, China*

[2]*Ecole Polytechnique Fédérale de Lausanne (EPFL), Swiss Plasma Center (SPC), CH-1015 Lausanne, Switzerland*

**Abstract**

The influence of charge trap states in the dielectric boundary material on capacitively coupled radio-frequency plasma discharge is investigated with theory and Particle-in-cell/Monte Carlo Collision simulation. It is found that the trap states of the wall material manipulated discharge properties mainly through the varying ion-induced secondary electron emission (SEE) coefficient in response to dynamic surface charges accumulated within solid boundary. A comprehensive SEE model considering surface charging is established first, which incorporates the valence band electron distribution, electron trap density, and charge trapping through Auger neutralization and de-excitation. Theoretical analysis is carried out to reveal the effects of trap states on sheath solution, stability, plasma density and temperature, particle and power balance, etc. The theoretical work is supported by simulation results, showing the reduction of the mean radio-frequency sheath potential as charging-dependent emission coefficient increases. As the gas pressure increases, a shift of the maximum ionization rate from the bulk plasma center to the plasma-sheath interface is observed, which is also influenced by the trap states of the electrode material where the shift happens at a lower pressure with traps considered. In addition, charge traps are proved helpful for creating asymmetric plasma discharges with geometrically symmetric structures, such effect is more pronounced in γ–mode discharges.

[1*] Email address: gjzhang@xjtu.edu.cn
[2*] Email: anbang.sun@xjtu.edu.cn

# I. Introduction

Capacitively coupled radio–frequency (RF) discharge plasmas are widely applied in various fields, such as surface processing, etching, film deposition and medicine [1-4]. Numerous studies in experiment, theory and simulation have been conducted to investigate the properties of capacitively coupled plasma (CCP) discharges in both concept and application [5-7]. In typical CCP discharges, ions are accelerated by mean RF sheath and collide with the solid surface, leading to the ion–induced secondary electron emission (ISEE). The electrons emitted from the boundary in turn induce ionization and modify plasma properties. Consequently, the plasma–surface interaction (PSI) plays a vital role in determining the properties of CCP discharges.

In previous works, it has been confirmed experimentally that the secondary electron emission (SEE) can significantly modify important plasma parameters, such as the temperature, density, and particle/power balance [5, 6]. Previous theoretical studies indicate that the SEE significantly modifies the plasma discharge characteristics [8, 9], these works also express the RF sheath as a function of the source amplitude and the SEE coefficient, giving quantitative current–voltage characteristic, sheath capacitance, conductance in the presence of boundary electron emission [10, 11]. In addition to experimental and theoretical approaches, numerical simulations provide further insight into the physical details of PSI in CCP discharges. The effect of boundary emission has been investigated with simulation and several simplifications of PSI are made to facilitate the modelling. In previous research, a constant ion induced secondary electron yield (ISEY) is typically set as the boundary condition and the impact of ISEE is studied [9, 12-20]. More recent works expand the scope of PSI to include the electron induced secondary electron emission (ESEE), ion/electron reflection, etc. The model proposed by Horváth *et al* includes the ESEE and electron backscatter, indicating that the ESEE strongly affects the plasma density and ionization dynamics [21]. Sun *et al* also provided a large number of studies regarding the effects of electron–surface interactions [22, 23]. In addition, as reported in previous works, asymmetric boundary conditions would also induce plasma asymmetry [9, 12, 24, 25].

In above numerical studies, the secondary electron emission yield (SEY) is regarded as a static function of the incident particle and solid boundary. For the ISEE, it is often set as a constant, which for slow ions is determined by the ion type and wall material. The energy–depended ISEE has also been investigated for heavy ions [19]. Whereas the ESEE only depends on the incoming electron energy for certain types of electrode [21]. Such simplifications ignore the solid–state physics nature of electron extraction from the boundary, while in reality dielectric materials (e.g. $SiO_2$) are frequently used in plasma processing applications and the emission coefficient could be dynamic in the course of a discharge due to the presence of trap states [26, 27]. Regarding the ISEY of a dielectric, there have been some previous works investigating the calculation of the ISEY. Motoyama *et al* discussed the relationship between the ISEY and dielectric energy band structure [28]. Moreover, Yoon *et al* studied the defect energy level's influence on the ISEY of uncharged dielectric surface and found the existence of a defect energy level which could improve the value of ISEY [29, 30]. Charge traps widely exist in dielectric materials as a result of chemical bonds breaking, the disorder of lattice, the impurity, bubbles, etc. [31-34] So unlike metals, charge traps play an active role for the ISEE from the dielectric due to the charge accumulation therein. When surface charges are trapped inside the boundary it is conceivable that the internal field will be shifted, therefore influencing the extraction of Auger electrons from the solid material. Such a process, although seemingly common, is currently ignored in most, if not all, CCP simulations, and its effects on the RF plasma properties remains unknown.

In this work, a 1D3V (spatially one–dimensional, three velocity coordinates) particle–in–cell / Monte Carlo collision (PIC/MCC) simulation is implemented to investigate the influence of trap states. In the ISEE model, the impact of charge traps is taken into account and an ISEY controlled by surface charges is calculated as presented in Section II. To facilitate more accurate simulation, a realistic experimentally obtained electron distribution is employed and two processes, i.e. Auger neutralization and de–excitation, are taken into account. The effect of charge traps on plasma properties, such as the sheath potential as well as particle and power balance, is theoretically studied and presented in Section III. In addition, the impact of charge–controlled ISEY on plasma sheath instability is also discussed, revealing why the I–V trace of the plasma sheath becomes more stable when charge traps in the electrode material are considered. In Section IV, the impact of charge traps on plasma properties, including the plasma density, sheath potential, heating rate of electrons and ions, ionization rate and electron mean energy are investigated to support theoretical predictions. The influence of gas pressure is considered as well. The sensitivity of plasma properties to the ISEY is studied from $p$ = 4 Pa to $p$ = 80 Pa and the effect of charge traps at different pressures is described and explained. Finally, the asymmetry induced by charge traps is discussed providing an option for creating asymmetric plasmas.

## II. Calculation of ISEY controlled by surface charges

It's widely believed that the ISEE is mainly triggered by Auger neutralization and Auger de–excitation for slow ions [35, 36]. The ISEY in these two processes is mainly determined by the distribution of the electrons as well as vacant states in the energy band, which can be controlled by the surface charge accumulation [35, 36]. In this section, the electrode structure as shown in Fig. 1 is used and the relationship between the ISEY and electrode surface charges is considered with a realistic electron distribution in the valence band, so that a more precise simulation can be set up. As shown in Fig. 1, the electrodes are perpendicular to the *x* axis, $L$ is the distance between electrodes, $w$ is the thickness of two electrodes and the electrode material is chosen as $SiO_2$. The selection of dielectric material does not influence the general conclusions obtained below. An alternating voltage with a typical frequency of 13.56 MHz is applied to the powered electrode, while the other electrode is grounded.

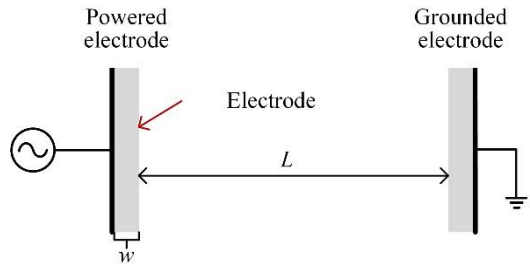

Fig. 1 Sketch of electrode structure

**A. Model description.**

As shown in Fig. 2, when ions approach the electrode surface, there are two possible electron extraction mechanisms: Auger neutralization and de–excitation [35, 36]. For Auger neutralization, as shown in Fig. 2 (a), the ion is neutralized with electron 1 from the electrode material, the energy released excites electron 2. Subsequently, electron 2 may escape and become a secondary electron (SE). While for Auger de–excitation, a resonance neutralization firstly occurs, forming an excited atom or molecule. As explained in Fig. 2 (b), excited electron 1 jumps down and the energy released excites electron 2; or in another case, electron 1 tunnels to the ground state and electron 2 absorbs the released energy, leading to a possible electron escape.

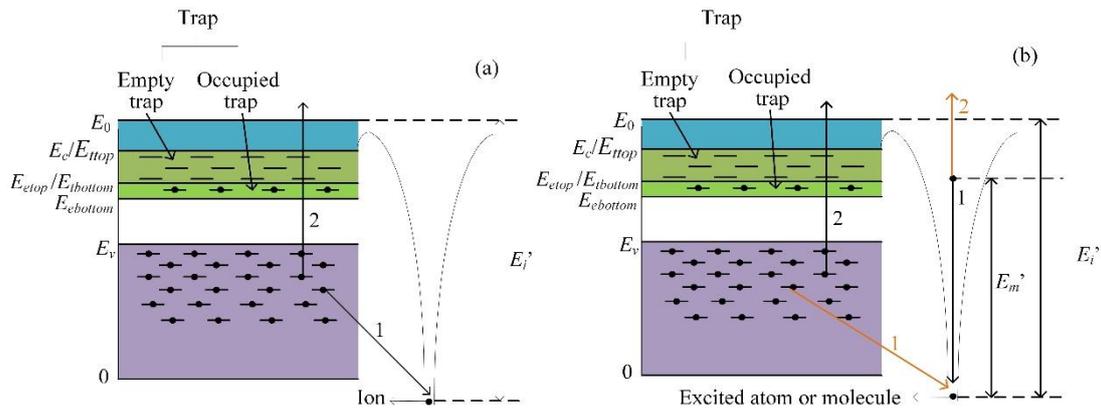

Fig. 2 Sketch of (a) Auger neutralization and (b) de–excitation process (Energy band structure, dielectric)

TABLE I Variable symbols

| Symbol | Description |
| --- | --- |
| $E_{etop}$ | Top energy of trapped electrons |
| $E_{ebottom}$ | Bottom energy of electron traps |
| $E_{ttop}$ | Top energy of electron traps |
| $E_{tbottom}$ | Bottom energy of empty traps |
| $E_0$ | Vacuum level |
| $E_v$ | Top energy of valence band |
| $E_c$ | Bottom energy of conduction band |
| $N_e$ | Trap density |
| $E_i{'}$ | Ionization energy at a distance $s_m$ from the solid surface |
| $E_m{'}$ | Excitation energy at a distance $s_m$ from the solid surface |

**B. Calculation of ISEY**

In this section, the relationship between the ISEY and electrode surface charge density is calculated and the meaning of some symbols are given in TABLE I. In order to obtain a more accurate calculation, a realistic electron distribution in the valence band, obtained by experiments rather than a constant value, is employed [37].

Firstly, the ISEY contributed by Auger neutralization is considered. In this process, trapped electrons are regarded as valence band electrons in the calculation. According to previous work, the distribution of excited electrons, $N_i^N$, can be calculated as follows [35, 36]:

$$N_i^N(E) \propto \rho_0(E) \cdot \int \int N(E_1)N(E_2)\delta(E_1 + E_2 + E_i{'} - E_0 - E)dE_1 dE_2 \tag{1}$$

where $\rho_0(E)$ refers to the distribution of the vacant state in the energy band as shown in Eq. (2); note that the distribution of empty states in the conduction band is proportional to $(E - E_c)^{1/2}$ for $E > E_c$ [38]; $N(E)$ is the electron distribution in the energy band given by Eq. (3), including trapped electrons and valence band electrons; $\delta$ is the Dirac delta function; as shown in Fig. 2 (a), $E_1$ and $E_2$ is the energies of electron 1 and 2, respectively; $E$ is the electron energy. Note that we use a simplification here for $E_i{'}$ and $E_m{'}$: since a transition occurs when an ion travels a long way to the solid surface, the energy $E_i{'}$ and $E_m{'}$ can be different depending on the distance from the ion to the solid surface when the transition happens. Thus, the desired ISEY $\gamma$ must be an average of ISEY $\gamma'$ over a distance from solid surface $s$. However, it is known from experimental observations that transitions occur collectively when the ions are at a certain distance from solid surface $s = s_m$ [35, 36]. Therefore, by substituting the energy at $s = s_m$, we can obtain a good approximation of $\gamma$.

$$\rho_0(E) = \begin{cases} 0 & 0 < E < E_{tbottom} \text{ or } E_{ttop} < E < E_c \\ N_e & E_{tbottom} < E < E_{ttop} \\ N_c\sqrt{E - E_c} & E > E_c \end{cases} \qquad (2)$$

$$N(E) = \begin{cases} N_v & 0 < E < E_v \\ N_e & E_{ebottom} < E < E_{etop} \\ 0 & E_v < E < E_{ebottom} \\ & \text{or } E > E_{etop} \end{cases} \qquad (3)$$

In order to simplify the expression, an operator $T^N$ is defined

$$T^N(E) = \begin{cases} \int_0^E N(E + \Delta) \cdot N(E - \Delta) d\Delta & 0 < E < \frac{E_{etop}}{2} \\ \int_0^{E_{etop}-E} N(E + \Delta) \cdot N(E - \Delta) d\Delta & E_{etop}/2 < E < E_{etop} \end{cases} \qquad (4)$$

Therefore, the distribution of excited electrons, $N_i^N$, can be written as

$$N_i^N(E) \propto \rho_0(E) \cdot T^N\left(\frac{E+E_0-E_{i'}}{2}\right) \qquad (5)$$

Knowing this, the distribution of escaped electrons can be calculated by

$$N_0^N(E) = P_e(E) \cdot N_i^N(E) \qquad (6)$$

where $P_e(E)$ represents the escaping probability controlled by the energy of excited electrons [38, 39]

$$P_e(E) = \begin{cases} 0 & E < E_0 \\ 0.5 * \dfrac{\left(1-\sqrt{\frac{(E_0-E_c)}{(E-E_c)}}\right)}{\left(1-0.967*\sqrt{\frac{(E_0-E_c)}{(E-E_c)}}\right)} & E > E_0 \end{cases} \qquad (7)$$

Finally, the ISEY triggered by Auger neutralization can be obtained as

$$\gamma^N = \int N_0^N(E) dE \,/\, \int N_i^N(E) dE \qquad (8)$$

Subsequently, the ISEY due to Auger de–excitation process is calculated. During this process, the distribution of excited electrons, $N_i^D$, is in the form of [35, 36]

$$N_i^D(E) \propto \rho_0(E) \cdot \int N(E_2) \delta(E_2 + E_m' - E) dE_2 \qquad (9)$$

where the variables have been defined above. Here, $T^D$ is chosen as

$$T^D(E) = N(E) \qquad (10)$$

Therefore $N_i^D(E)$ can be written as

$$N_i^D(E) \propto \rho_0(E) \cdot T^D(E - E_m') \qquad (11)$$

Similar to Auger neutralization, the energy distribution of escaped electrons can be calculated by

$$N_0^N(E) = P_e(E) \cdot N_i^N(E) \qquad (12)$$

In conclusion, the ISEY caused by Auger de–excitation is

$$\gamma^D = \int N_0^N(E) dE \,/\, \int N_i^D(E) dE \qquad (13)$$

In this work, SiO$_2$ is chosen as the electrode material and most quantities of interest are available in existing experiment data and previous numerical works. $N_v$ is obtained from experimental results as shown in Fig. 3 [37], $N_e = 5\times10^{22}$ eV$^{-1}$m$^{-3}$, $E_g = E_c - E_v = 9.2$ eV [40], $E_i' = 23.27$ eV [29] and $E_m' = 19.81$ eV [41]. Previous works also provided detailed calculations of $N_e$ and revealed its mechanism [42]. Here to facilitate later calculations in this paper, we employ the result from Yao *et al* [39]. Some functions used in Eq. (1) - (12) are presented in Fig. 3, such as the distribution of vacant states in the energy band $\rho_0(E)$, the electron distribution in the energy band $N(E)$ and the escaping probability $P_e(E)$. The calculation results during Auger Neutralization and De–excitation, including the operator $T^D$ and $T^N$, the distribution of excited electrons $N_i^D$ and $N_i^N$, the energy distribution of escaped electrons $N_i^D$ and $N_i^N$, are given as well to help understanding, where an uncharged electrode is considered.

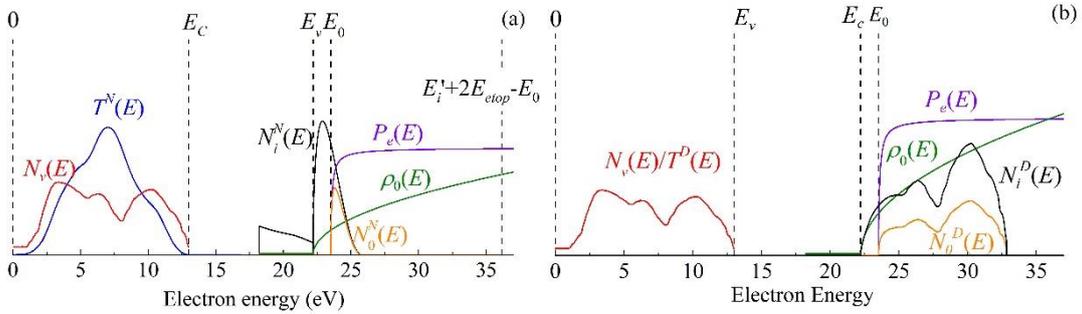

Fig. 3 Example of calculation results for (a) Auger Neutralization and (b) Auger De–excitation

In the end, the total ISEY can be calculated and expressed as

$$\gamma = P_N\gamma^N + P_D\gamma^D \qquad (14)$$

assuming that $P_D = 1/2R_e$ and $P_N = 1 - P_D$, where $R_e$ represents the density ratio of electron traps to the total electron density in the energy band [28]. Note that the density of electron traps is much smaller than valence band, thus the ISEY is mainly determined by the process of Auger neutralization.

From our calculations it can be shown that when charges accumulate on the electrode surface, electrons occupy the traps in the energy band, leading to the change of the electron/vacant state distribution, and finally affecting the ISEY. If we assume that $dE = E_{etop} - E_{ebottom}$ corresponds to the occupied traps, the surface charge density then can be obtained as $\sigma = edEN_el$, where $l$ is the penetration depth and $e$ is the elementary charge. It's usually assumed that electrons are deposited in a single–atom–layer[39], hence, $l$ is set as $10^{-11}$ m in this paper. This value can vary with the material type, and may influence the effects of charge accumulation. However, the choice of this value won't change the general tendency presented in the manuscript, thus it does not seem to impact the relevant conclusions significantly.

The calculated curves of the ISEY–surface charge density are given in Fig. 4. With the accumulation of negative charge, as the ISEY for both aforementioned processes changes, the ISEY of Auger neutralization increases more sharply. When the charges accumulated in the electrode occupy all the traps, the ISEY stays constant. The impact of surface charge accumulation is mainly due to the modification of the electron/vacant state distribution.

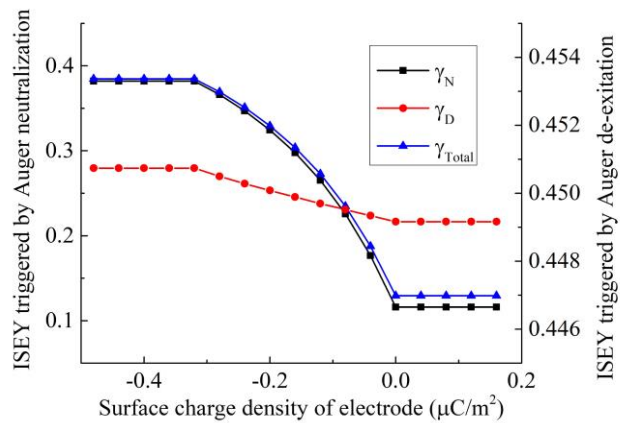

Fig. 4 ISEY of charged surface

**III. Theoretical analyses**

In this section, the impact of charge traps on plasma discharge properties, including the influence on sheath potential, particle and power balance as well as plasma sheath instabilities, is analyzed theoretically. Some of the symbols used below are presented in TABLE II.

TABLE II Variable symbols

| | |
|---|---|
| $n_e$ | Total electron density in the plasma sheath and pre–sheath |
| $n_i$ | Ion density in the plasma sheath and pre–sheath |
| $n_0$ | Plasma density at the sheath edge |
| $n_{ep}$ | Plasma electron density in the plasma sheath |
| $n_{se}$ | Secondary electron density in the plasma sheath |
| $n_{ep0}$ | Plasma electron density at the sheath edge |
| $n_{se0}$ | Secondary electron density at the sheath edge |
| $n_{sew}$ | Secondary electron density at wall |
| $\varepsilon_{i0}$ | Ion energy at the sheath edge |
| $v_i$ | Ion velocity in the plasma sheath |
| $v_{i0}$ | Ion velocity at the sheath edge |
| $v_{se}$ | Secondary electron velocity in the plasma sheath |
| $v_{sew}$ | Secondary electron velocity at wall |
| $m_e$ | Electron mass |
| $m_i$ | Ion mass |
| $\varphi$ | Electric potential in the plasma sheath |
| $\varphi_w$ | Electric potential at wall |
| $\omega$ | Frequency of applied voltage |
| $\omega_e$ | Electron plasma frequency |
| $\omega_i$ | Ion plasma frequency |
| $T_{ep}$ | Temperature of plasma electrons |
| $T_{ip}$ | Temperature of plasma ions |
| $\gamma$ | ISEY |

To begin with, a brief discussion of the plasma sheath in the presence of trap states is in order. The sheath solution of emissive boundary was originally proposed by Hobbs and Wesson[43]. The approach assumed a constant electron induced secondary electron emission yield (ESEY). The Bohm criteria, charge neutrality and flux balance were used to derive the sheath potential for an electron–emitting boundary. In the case of the RF sheath in the CCP discharge, the real–time plasma sheath becomes a superposition of the mean RF sheath and an oscillating component [44]. The ions respond to the mean space potential whereas electrons respond to the real–time space potential, since in a typical CCP discharge the relationship $\omega_i < \omega < \omega_e$ is satisfied, where $\omega$ is the frequency of applied voltage, $\omega_e$ is the ion plasma frequency and $\omega_e$ is the electron plasma frequency. Thus, the mean RF sheath determines the ion incident energy at the solid surface, and is therefore crucial in view of the numerous applications related to CCP [1]. Furthermore, the mean RF sheath potential is a function of the applied voltage amplitude and the emission coefficient, but simultaneously considering the two factors leaves no analytical solution to the RF sheath [10]. Instead, a qualitative analysis using an example of a floating boundary is given below to illustrate the influence of trap states on sheath properties. The general trend of the RF sheath in CCP discharges should be analogous.

In the following considerations, the potential of the sheath edge is assumed to be 0. Note that the sheath edge connects the presheath and sheath, where the entering speed of ions is characterized by Bohm criterion. It is also defined as the point outside of which the charge neutrality breaks down, so here we take $n_e = n_i = n_0$ and $\varphi = 0$ at the sheath edge. Rigorously speaking, the ion velocity distribution function $f_i$ at the plasma sheath edge can be found from the Boltzmann equation [45]:

$$v \frac{\partial f_i}{\partial x} + \frac{e}{m_i} E(x) \frac{\partial f_i}{\partial v} = -\frac{|v|}{\lambda_{CX}} f_i + \delta(v) Q(v) \tag{15}$$

where the ion charge exchange collision and ionization constitute the source terms, $v$ is the ion velocity, $E(x)$ is the self–consistent electric field, $\lambda_{CX}$ is the charge exchange mean free path, $\delta$ is the Dirac delta function. $Q$ is the rate by which ions are produced per volume at zero velocity which equals to

$$Q(v) = \int \frac{|v|}{\lambda_{CX}} f_i dv + v_{iz} n_e \tag{16}$$

where $v_{iz}$ is the ionization rate and $n_e$ is the electron density. $Q(v)$ could be further expressed by Wannier operator [46]. Solving Eq. (16) with ionization–free assumption in presheath gives the ion velocity distribution function at the sheath edge, which can be written as $f_{i0} = \eta n_0 \sqrt{\frac{2m_i}{T_{ep}}} K'(-\frac{m_e v^2}{2T_{ep}}) \exp[K(-\frac{m_e v^2}{2T_{ep}})]$, with $n_0$ the plasma density at the sheath edge, $\eta$ the eigenvalue, and $K(x) \approx (0.185 - 0.011x)(1 - 2x - e^{-2x})$ [47], with which the ion flux at the sheath edge is calculated to be:

$$\Gamma_i = \int_0^\infty v f_{i0}(v) dv = \eta n_0 \sqrt{\frac{2T_{ep}}{m_i}} \tag{17}$$

The general Bohm criterion is dictated by $\eta \geq 2^{-0.5}$. Riemann's calculation gives $\eta = 0.88161$ but above solution is only valid when no SE exists in sheath [47]. In the following deductions, we adopt the cold ion assumption ($T_{ip} \ll T_{ep}$), which is less accurate than above but can provide solvable equations.

The energy and flux conservation of ions in the plasma sheath can be expressed as .

$$\frac{1}{2} m_i v_i^2 - \varepsilon_{i0} = -e\varphi \tag{18}$$

$$n_0 v_{i0} = n_i v_i \tag{19}$$

where $\varepsilon_{i0} = 1/2 m_i v_{i0}^2$ is the monoenergetic ion energy. Thus, the ion density in the plasma sheath is simply given by

$$n_i = n_0 \left(1 - \frac{e\varphi}{\varepsilon_{i0}}\right)^{-1/2} \tag{20}$$

For SEs, the velocity distribution function can be considered in a similar fashion as show in Qing's work [48]. Note that here the temperature of SEs ($T_{se}$) approaches 0, thus we can utilize energy conservation, flux conservation and secondary emission coefficient given respectively in Eq. (21) - (23).

$$\frac{1}{2} m_e v_{se}^2 = e(\varphi - \varphi_w) \tag{21}$$

$$n_{sew} v_{sew} = n_{se} v_{se} \tag{22}$$

$$n_{sew} v_{sew} = \gamma \cdot n_0 v_{i0} \tag{23}$$

Combining above equations, the secondary electron density in the plasma sheath is then given by

$$n_{se} = \gamma n_0 \sqrt{\frac{\varepsilon_{i0}}{\mu e(\varphi-\varphi_w)}} \quad (24)$$

Subsequently, the generalized Bohm criterion presented in Eq. (26) is considered. $n_i$ and $n_{se}$ have been given in Eq. (20) and (24), respectively, and $n_{ep}$ is given by Eq. (25).

$$n_{ep} = n_{ep0}\exp(\frac{e\varphi}{T_{ep}}) = (n_0 - n_{se0})\exp(\frac{e\varphi}{T_{ep}}) \quad (25)$$

$$\frac{\partial n_i}{\partial \varphi} - \frac{\partial (n_{ep}+n_{se})}{\partial \varphi}\bigg|_{\varphi=0} \leq 0 \quad (26)$$

Neglecting the terms containing $1/\mu$, the marginal solution of Eq. (26) is calculated to be $\varepsilon_{i0} \approx 1/2 T_{ep}$.

Note that one omission here is that the energetic plasma electron penetrating the sheath will not return. As a result, the high–energy part of the electron velocity distribution function is strongly depleted and often termed as the loss cone in the velocity space. Here the loss cone can be discarded because its influence becomes diminishes at a high sheath potential [49], which is the case in the CCP discharge.

Finally, solving the flux balance Eq. (27) at the wall

$$\Gamma_{ep} - \gamma\Gamma_i = \Gamma_i \quad (27)$$

The sheath potential $\varphi_w$ can be expressed as a function of ISEY $\gamma$

$$\ln(\sqrt{\frac{\mu}{2\pi}}\frac{1}{1+\gamma}) = -\frac{e\varphi_w}{T_{ep}} \quad (28)$$

Clearly, the introduction of the ISEE lowers the sheath potential. In Section II we have shown that $\gamma$ is impacted by the electrode surface charge density, thus the relationship between the surface charge density and sheath potential, considering different trap densities in the electrode material, can be obtained by combining Eq. (14) and (28). The results for different trap densities are shown in Fig. 5 bellow.

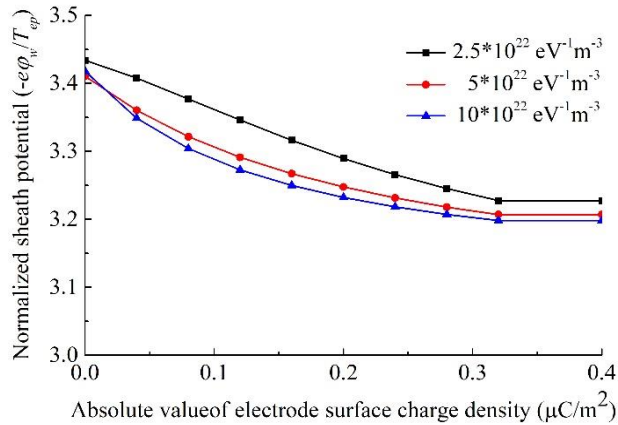

Fig. 5 Electric potential with different trap densities of electrode material.

Furthermore, we would like to analyze the particle and power balance in the presence of charge traps. Assuming a collisionless sheath, the plasma loss at the boundary must be compensated by the ionization in the bulk plasma.

$$V_{dis}K_{iz}n_p n_g = n_p \sqrt{\frac{T_{ep}}{m_i}} A_{dis} \tag{29}$$

where $V_{dis}$ is the discharge volume; $K_{iz}$ is the ionization rate; $A_{dis}$ is the effective area of discharge; while $n_p$ and $n_g$ are the densities of plasma and gas atom, respectively. It can be found that the particle balance is not greatly changed by the SEE, as the ion and electron flux (plasma electron flux minus SE flux) at the wall remain well balanced,

$$\Gamma_{ep} - \Gamma_{se} = \Gamma_i \tag{30}$$

We can therefore state that charge traps only have a limited influence on particle balance if the ionization induced by SEs is ignored. Note that the presence of SEE can increase the plasma density, especially in γ–mode discharges where the ionization due to the electrons emitted from the boundary becomes non–negligible, which would inevitably alter the ion flux in the sheath.

In addition, the impact of charge traps on power balance is considered. It's known that the power balance in the CCP discharge can be described by

$$P_{RF} = P_{loss} \tag{31}$$

where $P_{RF}$ is the total input power from source and $P_{loss}$ is the total power consumption in the discharge area. $P_{RF}$ can be expressed as Eq. (32)

$$P_{RF} = P_{ohmic} + P_{stoc} \tag{32}$$

where $P_{ohmic}$ is from ohmic heating and $P_{stoc}$ is from stochastic heating.

$P_{loss}$ can be separated into four parts: the power loss at the boundary of plasma electrons $P_{edge,ep}$ and ions $P_{edge,i}$, the power loss due to SE $P_{se}$ and the power loss due to inter–particle collisions $P_{coll}$

$$P_{loss} = P_{edge,i} + P_{edge,ep} + P_{se} + P_{coll} \tag{33}$$

The expressions of the power losses on the right–hand side (RHS) of Eq. (33) are given by [8]

$$P_{edge,i} = 2\Gamma_i \varepsilon_i \qquad \text{with } \varepsilon_i = 2T_i + \frac{T_{ep}}{2} + e\overline{\varphi_{sh}} \tag{34}$$

$$P_{edge,ep} = 2\Gamma_{ep}\varepsilon_{ep} \qquad \text{with } \varepsilon_{ep} = 2T_{ep} \tag{35}$$

$$P_{se} = 2\Gamma_{se}\varepsilon_{se} \qquad \text{with } \varepsilon_{se} = \varepsilon_i \tag{36}$$

It's obvious that the charge traps mainly influence $P_{se}$, which can be seen as the power needed to accelerate the SEs due to the sheath field. In the sheath, SEs are accelerated by the sheath electric field and serves as a power drain. Moreover, the SEE from the boundary could increase the plasma density in certain cases, thus augmenting the ion flux and enhancing the power loss due to the escaping ions. To sum up, the impact of charge traps on power loss can be explained by the modification of electron emission at the boundary.

The final point of discussion is the plasma sheath instability. It is widely known that the presence of SEE at the boundary can change the plasma sheath instability [50]. The boundary emission alters the flux balance and therefore changes the response V–I characteristics of the sheath [51]. Intense boundary emission leads to quasiperiodic relaxation oscillations switching the sheath between stable and unstable regimes [52]. The instability of the plasma sheath can be understood by the fact that a positive perturbation of the sheath potential $\varphi_{sh}$ (more negative charges stored in boundary) must be compensated by a reduction of the electron flux such that the fluctuation decays[50, 51]. This can be expressed by

$$\frac{\partial \Gamma_e}{\partial \varphi_{sh}} < 0 \tag{37}$$

where $\varphi_{sh} = |\varphi_w|$ is the sheath potential (positive) and $\Gamma_e = \Gamma_{ep} - \Gamma_{se}$ is the net electron influx toward the boundary, where $\Gamma_{ep}$ is the electron flux from the bulk plasma and $\Gamma_{se}$ is the one emitted from the boundary due to the ISEE.

When the boundary electron emission is neglected, the net electron influx can be expressed as $\Gamma_e = \Gamma_{e,edge} \exp(-e\varphi_{sh}/T_e)$, where $\Gamma_{e,edge}$ is the electron influx at the sheath edge which depends on the plasma density and temperature. Eq. (37) is valid for absorbing boundaries whereas for an emission boundary, the relationship is more complex.

With a floating boundary, $\Gamma_e = \Gamma_i$ is always true due to charge conservation. While under the RF voltage, this equation is only valid for the time–averaged flux. This simplified relation is employed to obtain analytical expression for cross comparison. According to Eq. (27), there is $\Gamma_e = \Gamma_i = \left(\frac{1}{1+\gamma}\right)\Gamma_{ep}$, Eq. (37) thus can be rewritten as

$$\frac{\partial \Gamma_e}{\partial \varphi_{sh}} = \left(\frac{1}{1+\gamma}\right)\frac{\partial \Gamma_{ep}}{\partial \varphi_{sh}} - \frac{1}{(1+\gamma)^2}\frac{\partial \gamma}{\partial \varphi_{sh}}\Gamma_{ep} < 0 \tag{38}$$

The RHS of Eq.(38) contains two terms. $\frac{\partial \Gamma_{ep}}{\partial \varphi_{sh}} < 0$ is true for classic Debye sheath where the wall potential is well below the plasma potential because the sheath potential prevents the electrons from moving toward the boundary. It is not the case for the inverse sheath where the wall potential is higher than the plasma potential. Simple flux balance analysis shows that the necessary condition for the inverse sheath cannot be achieved when the ISEE triggers the boundary emission [11]. Then the first term is negative since $\gamma > 0$ and $\frac{\partial \Gamma_{ep}}{\partial \varphi_{sh}} < 0$, thus stabilizing the plasma sheath. Unlike the first term, the second term can change with charge traps. The question to be investigated is whether charge trapping stabilizes or destabilizes the plasma sheath, and what is the significance of such factor.

As shown in Section II, $\gamma$ is constant and independent of the sheath potential if the impact of traps is allowed to be neglected. Therefore, the second term is always zero with $\frac{\partial \gamma}{\partial \varphi_{sh}} = 0$. In this case, the ISEE does not impact the plasma instabilities.

However, if charge traps are considered, as shown above, $\gamma$ becomes a function of the charge density. The second term in Eq. (38) then can be written as

$$\frac{1}{(1+\gamma)^2} \frac{\partial \gamma}{\partial \varphi_{sh}} \Gamma_{ep} = \frac{\Gamma_{ep}}{(1+\gamma)^2} \frac{\partial \gamma}{\partial n_T} \frac{\partial n_T}{\partial \varphi_{sh}} \tag{39}$$

where $n_T$ is the density of trapped charges in the dielectric (take absolute value). It's obvious that $\frac{\Gamma_{ep}}{(1+\gamma)^2} > 0$, so that the value of the second term depends on $\frac{\partial \gamma}{\partial n_T} \frac{\partial n_T}{\partial \varphi_{sh}}$. From Section II, it's known that the ISEY increases as more negative charges accumulate in the dielectric material, thus $\frac{\partial \gamma}{\partial n_T} > 0$. The partial derivative $\frac{\partial n_T}{\partial \varphi_{sh}}$ is positive as well since the amount of charge trapped in the dielectric is believed to be equal to the positive charge in sheath (same amount, opposite sign) in the equilibrium state [53]. Furthermore, more charges in the sheath leads to a higher sheath potential. In conclusion, it can be found that the second term of Eq. (38) is positive when the charge traps are considered and that the traps are helpful for stabilizing the plasma sheath. The trap density is greatly related to the type of material, crystal form, manufacturing process, temperature and so on [31, 33, 54]. For applications where instabilities are unwanted, material with a higher trap density may be used to naturally stabilize plasmas.

## IV. Simulation Setup and Results

In this part, a simulation model is introduced to reveal the impact of charge traps on CCP discharges and support the theoretical predictions.

**A. Simulation setup**

In this work, a 1D3V CCRF discharge is considered with PIC/MCC simulation. For more detail of code validity, convergence and benchmark we refer to author's previous work [55]. The general findings for the convergence behavior reported in [56] were found to be valid for the present analysis as well. As shown in Fig. 1, the electrodes are perpendicular to the $x$ axis, the distance between the two electrodes is 6.7 cm, the thickness of the electrodes is 0.2 cm and the electrode material is chosen as $SiO_2$, relevant parameters and the calculation of ISEY are presented in Section II. The spatial step size is 0.1 mm with 711 grid points. An alternating voltage is applied to the powered electrode with a frequency of $f$ = 13.56

MHz and peak voltage of 210 V, while the other electrode is grounded. In this paper, the ESEE as well as electron reflection is neglected. This is because the electrons in our simulation have a low energy, leading to a quite small ESEY. Moreover, in the pressure range 4 to 80 Pa and applied voltage of 210V discussed in the text, the ESEE or reflection of electrons don't significantly impact the discharge properties according to previous research [12, 21]. However, at a lower pressure and higher applied voltage, 0.5 Pa and a voltage of 1000 V across 6.7cm for example, their impact could be considerable [21]. In addition, our aim here is to present the impact of taking into account the electron traps on the calculated discharge characteristics.

The initial conditions of the particles are as follows. Helium gas is set as the background gas, its temperature is fixed at 300 K and the pressure is held constant. The constant pressure values chosen for the simulations were $p$ = 4, 20, 40, 80 Pa. The collision processes considered in this simulation are shown in TABLE III. The cross sections (the excitation, ionization and elastic collision) are obtained from our previous work [57-59]. The initial particle weight (the number of real particles that a super–particle represents) is set as 20 and automatically updates during the simulation, for example, the number of macro particles is about 2000 at 4Pa and 45000 at 80Pa. The time step is fixed as d$t$ = 0.37×10$^{-11}$ s throughout the simulation process. The data is collected until the simulation reaches a periodic steady state. The computations performed took approximately 15 human hours for 4 Pa and 60 human hours for 80 Pa using 22 CPU cores.

TABLE III Collision processes considered in the simulation

| Reaction | Type | Energy threshold (eV) | References |
| --- | --- | --- | --- |
| He + $e$ → He + $e$ | Elastic collision | — | [57, 59] |
| He + $e$ → He* + $e$ | Excitation (triplet) | 19.82 | [57, 59] |
| He + $e$ → He** + $e$ | Excitation (singlet) | 20.61 | [57, 59] |
| He + $e$ → He$^+$ + 2$e$ | Ionization | 24.59 | [57, 59] |
| He + He$^+$ → He$^+$ + He | Elastic (backward) | — | [57, 58] |
| He + He$^+$ → He + He$^+$ | Elastic (isotropic) | — | [57, 58] |

**B. Simulation result**

In this section, the effect of charge traps on diverse plasma properties is analyzed with different assumptions for the boundary conditions at the powered and the grounded electrode. Four ISEY models are implemented at the electrodes: 1) $\gamma$ = 0 which is a complete absorbing boundary; 2) $\gamma$ = 0.116, which is the ISEY when the electrode is uncharged; 3) $\gamma$ = 0.382, which is the ISEY when the electrode is fully charged; and 4) an ISEY controlled by surface charges. Note that, the ESEE and electron reflection are neglected in all cases.

The results of the 1D3V simulation at gas pressures of 4Pa as well as 80Pa are shown in Fig. 6 and 7, respectively. Here, the time–averaged spatial distributions of the a) ion density $n_i$, b) electric potential, c) mean electron energy, d) heating rate of electrons $P_e$, e) ion power density heating rate of ions, and f) ionization rate is presented. The sheath edge is displayed as well, with the edge setting at the location where $(n_i - n_e)/n_i = 0.05$.

As shown in Fig. 6 (a) and 7 (a), the introduction of charge traps increases the plasma density. Clearly, the curve of the charge–controlled ISEY is always between those of $\gamma = 0.116$ (no charge accumulation) and $\gamma = 0.382$ (fully charged), indicating that the charge–controlled ISEY is greater than the ISEY of uncharged surface and smaller than that of a fully charged surface. Apparently, the ISEY of the electrode is augmented with the accumulation of negative charges as shown in Fig. 4 and it is always smaller than $\gamma = 0.382$. This statement can be confirmed by the results in Fig. 9, i.e. the electrodes are not fully charged (smaller than 0.32 μC/m$^2$) during the simulation. The change of the background gas pressure also makes a difference. The variation of the plasma density induced by charge traps is more significant at a high pressure ($p = 80$ Pa) and less remarkable at a relatively low pressure ($p = 4$ Pa). This can be explained by the increase of the ion flux toward the boundary. As discussed later and shown in Fig. 9, the ion flux toward the boundary is positively related with the gas pressure. Thus, the ion flux becomes much larger at the pressure of $p = 80$Pa due to the increase of plasma density, causing more SEs and generating further ions as well as electrons due to the ionization process compared with the situation at a lower pressure. As a result, the effect of charge traps becomes stronger.

In addition, the electric potential slightly decreases when charge traps are employed as shown in Fig. 6 (b) and 7 (b), in agreement with the conclusion presented in Section III. When charge traps are considered, the ISEY increases as discussed in Section II and shown in Fig. 4, resulting in a decrease of sheath potential as presented in Eq. (28) and Fig. 5. In contrast, the electrical potential is not influenced a lot by gas pressure, which is consistent with Eq. (28).

As shown in Fig. 6 (c) and (d) as well as Fig. 7 (c) and (d), the introduction of charge traps increases the heating rate of both electrons and ions in the plasma sheath. As discussed in Section II, charge traps augment the ISEY of electrodes. The influence of ISEY on heating rate has been discussed in Section III: a higher ISEY brings additional acceleration (i.e. the acceleration of SEs), resulting in a power drain and increasing the heating rate of electrons. Moreover, the plasma density is augmented by charge traps as explained above, consequently rising the electron flux toward the boundary and power loss due to electron escaping. While for the ions, the change of the heating rate in the plasma sheath is mainly caused by the increase of the plasma density and ion flux toward the boundary. The gas pressure also influences the effect of charge traps. As explained above, the impact of charge traps is more obvious at a higher pressure due to a larger flux toward the boundary and more SEs generated by the ISEE.

[Fig. 6 (e)](#) and [Fig. 7 (e)](#) present the time–averaged spatial distribution of the ionization rate. It appears that the ionization rate increases with the presence of charge traps at all pressures, however its effect varies at different pressures. At $p = 4$ Pa, the charge traps enhance the ionization in the center of the bulk plasma, while at $p = 80$ Pa, the ionization rate near the sheath–plasma interface increases. This is because the plasma sheath is approximately collisionless at low pressures, hence the SEs can easily reach the bulk plasma where they induce ionization. Nevertheless, at $p = 80$ Pa, electrons collide with gas atoms before reaching the center of the bulk plasma due to the dense background gas, creating ionization mainly near the sheath. In addition, the gas pressure considerably influences the distribution of the ionization as shown in [Fig. 8](#). It can be observed that the ionization mainly happens in the center of the discharge domain at low pressures, while the maximum ionization rate shifts closer to the sheath–plasma interface as the background pressure increases. This phenomenon can be explained by the decrease of the mean free path. At a higher pressure, the mean free path of electrons decreases due to the dense background gas, thus the electrons dissipate their energy obtained from the sheath within a smaller distance, creating a peak of the ionization rate near the sheath–plasma interface. It should be mentioned that charge traps also play a role in the shift of the maximum ionization rate, which will be discussed later.

The mean electron energies at different pressures are displayed in [Fig. 6 (f)](#) and [Fig. 7 (f)](#). Both figures demonstrate that the inclusion of charge traps increases the mean electron energy in the plasma sheath and decreases the one in the bulk plasma at all pressures. The electric potential does not change a lot as shown in [Fig. 6 (b)](#) and [Fig. 7 (b).](#) More SEs are generated when charge traps are considered and these SEs become energetic after they are accelerated in the sheath. As a result, the mean electron energy in the plasma sheath changes with the introduction of charge traps. Nevertheless, the presence of charge traps decreases the mean electron energy in the bulk plasma, because the SEs dissipate most of their energy near the sheath edge due to collision. The background gas pressure also influences the spatial distribution of the mean electron energy. At a low pressure ($p$ = 4 Pa), the electron mean energy increases in the plasma sheath and is somewhat uniform in the bulk plasma due to the acceleration by the electric field in the sheath. While at a higher pressure ($p$ = 80 Pa) there exists the maximum mean electron energy locating near the boundary. The main reason is that the SEs are accelerated by electric field in the bulk plasma, but they will then collide with gas atoms and lose energy, generating a peak near the sheath edge. The shift of the mean electron energy peak indicates collisions at the bulk–sheath interface and a peak shift of the ionization rate.

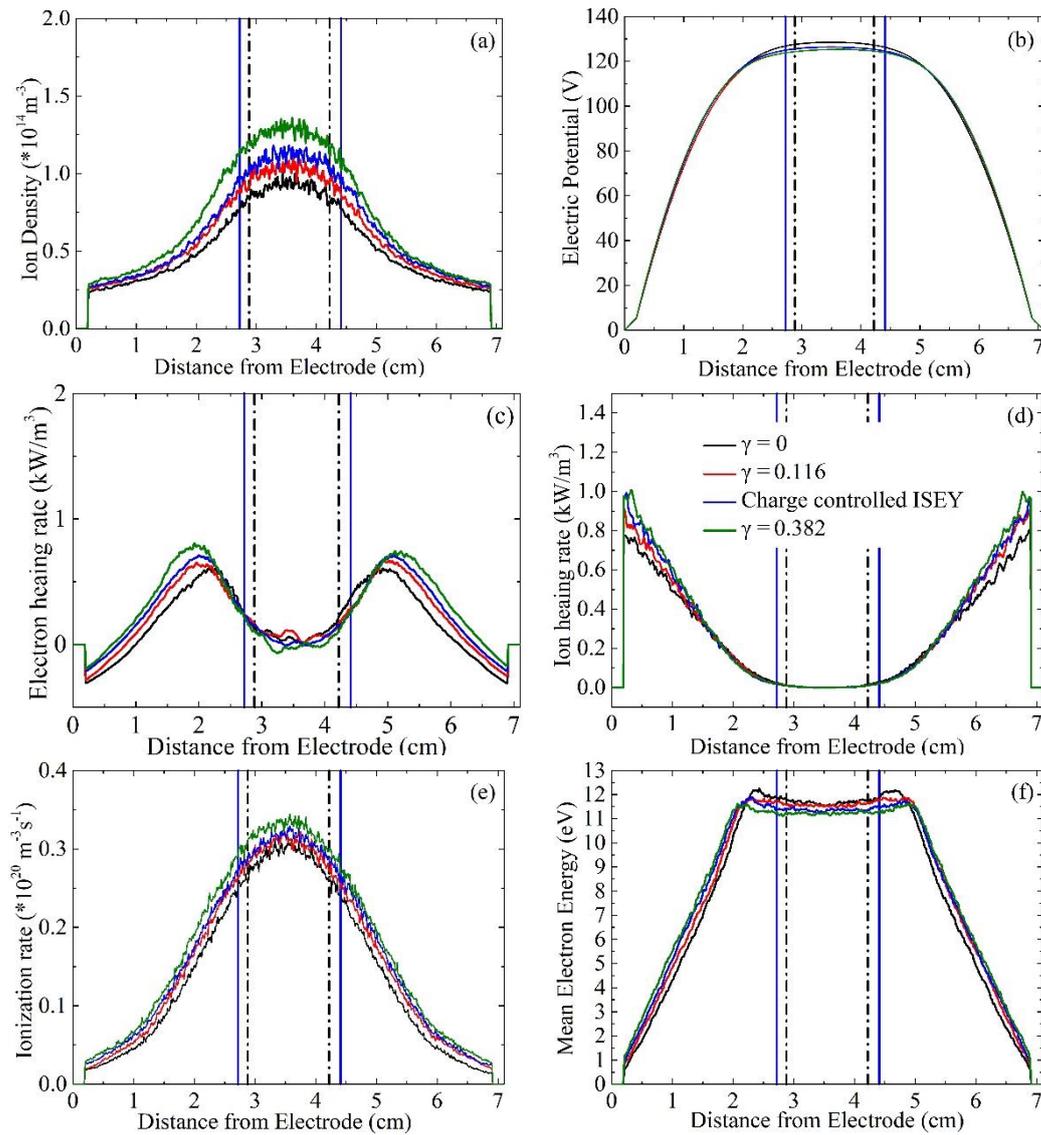

Fig. 6 Time–averaged spatial distribution of the a) ion density $n_i$, b) electric potential, c) mean electron energy, d) heating rate of electrons $P_e$, e) heating rate of ions $P_i$ and f) ionization rate at 4 Pa for four boundary condition assumptions: 1) $\gamma$ = 0 (completely absorbing boundary), 2) $\gamma$ = 0.116 (uncharged electrode surface), 3) $\gamma$ = 0.382 (fully charged electrode surface) and 4) charge–controlled ISEY.

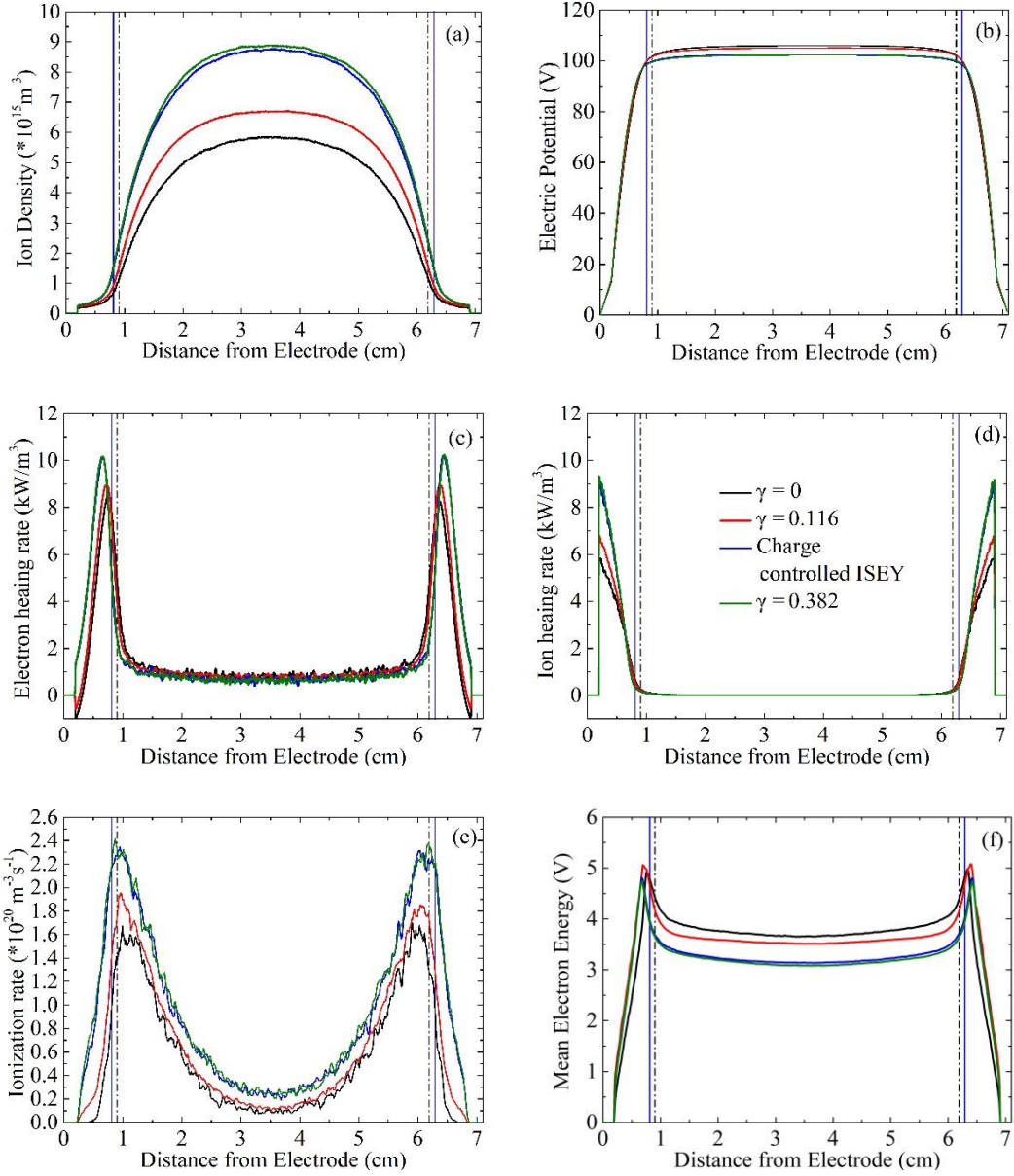

Fig. 7 Time–averaged spatial distribution of a) ion density $n_i$, b) electric potential, c) mean electron energy, d) heating rate of electrons $P_e$, e) heating rate of ions $P_i$ and f) ionization rate at 80 Pa for four boundary condition assumptions: 1) $\gamma = 0$ (completely absorbing boundary), 2) $\gamma = 0.116$ (uncharged electrode surface), 3) $\gamma = 0.382$ (fully charged electrode surface) and 4) charge–controlled ISEY.

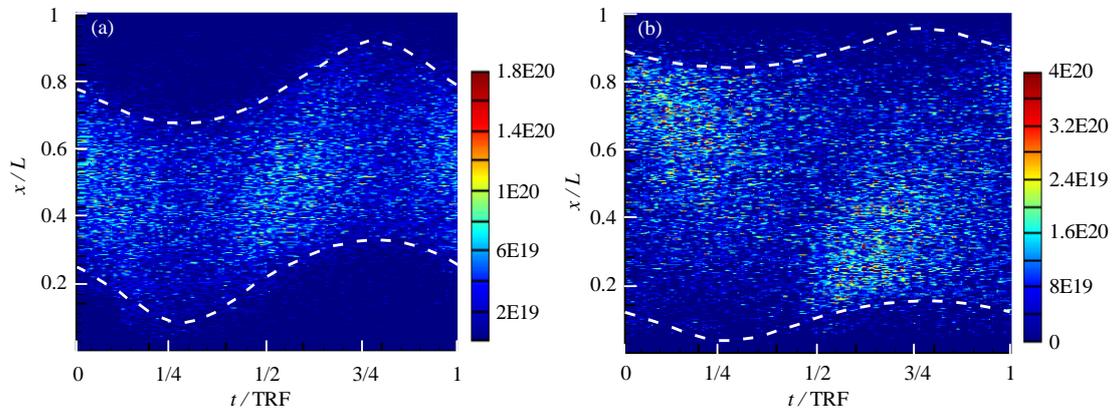

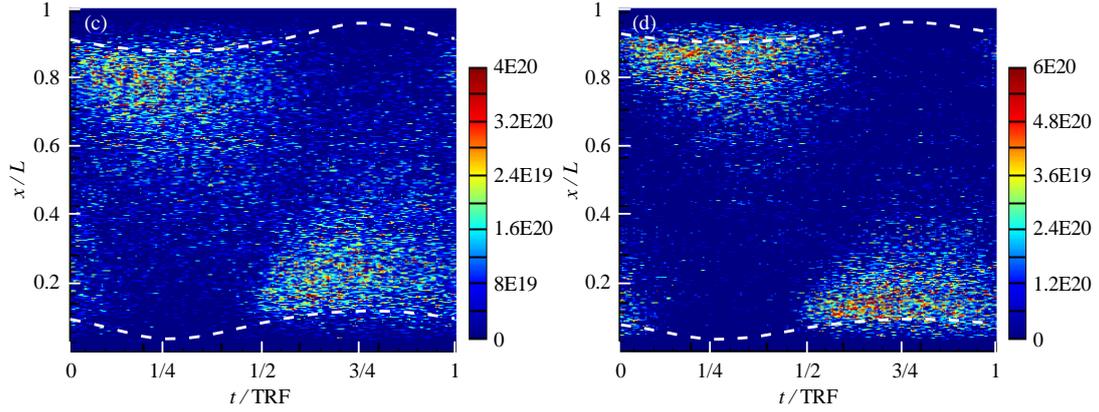

Fig. 8 Time–Spatial distribution of the ionization rate at a) $p$ = 4 Pa, b) $p$ = 20 Pa, c) $p$ = 40 Pa and d) $p$ = 80 Pa

Some important properties are shown as a function of gas pressure in Fig. 9, including the (a) peak ion density, (b) ion flux toward the wall, (c) surface charge density at electrode and (d) mean electron energy. The trends of these properties, like the ion density (Fig. 9 (a)) and ion flux (Fig. 9 (b)), have been discussed above. Note that with the gas pressure rising, the difference between the curve of charge–controlled ISEY and the one of $\gamma$ = 0.382 becomes bigger at first and then decreases. Such phenomenon is caused by the competition between two factors. On one hand, the impact of ISEY becomes more significant at a higher pressure, leading to a larger difference between the two aforementioned curves. On the other hand, the surface charge density increases when the pressure goes up as shown in Fig. 9 (d), hence the charge–controlled ISEY actually grows, making the two curves closer. Meanwhile, the effect of gas pressure is related to the ISEY model employed. When charge traps are considered, the impact of gas pressure becomes greater. Additionally, as mentioned before, Fig. 9 (c) indicates that the mean electron energy in the center declines as the gas pressure increases, which is consistent with the results shown in Fig. 6 (f) and Fig. 7 (f). In the end, there is a positive correlation between the surface charge accumulation and the background gas pressure as shown in Fig. 9 (d). It seems natural as denser background gas provides more charges, hence influencing the surface charge accumulation.

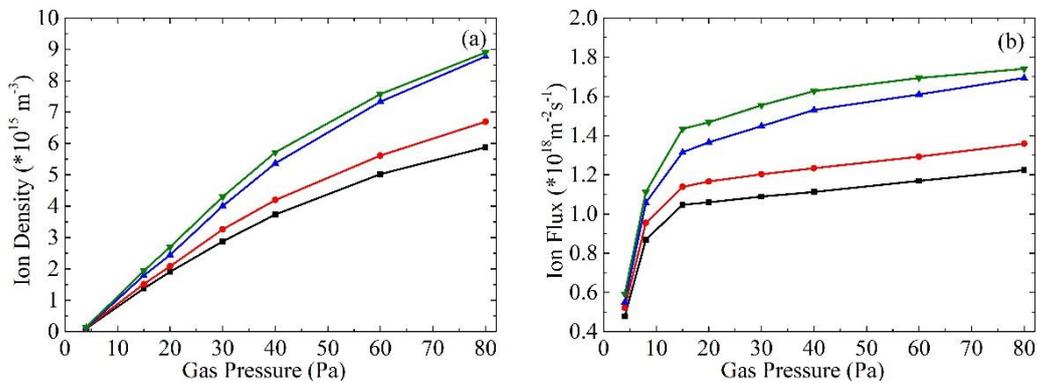

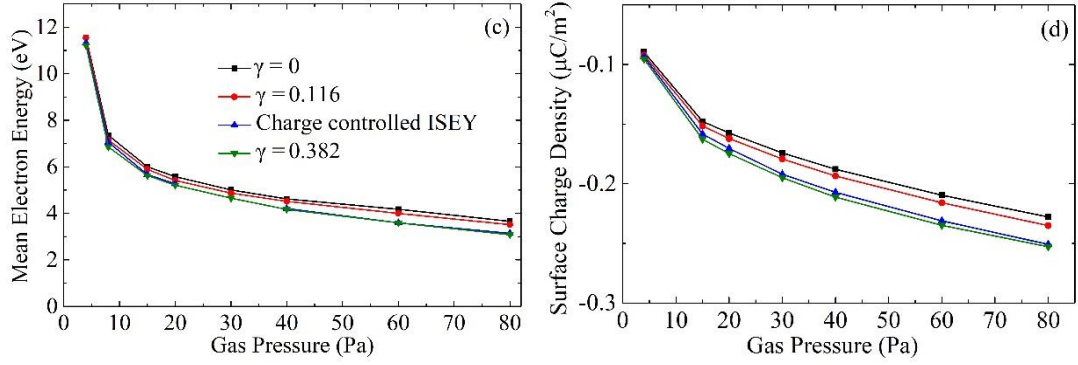

Fig. 9 Time–averaged (a) peak ion density, (b) surface charge density at electrode, (c) mean electron energy in the center and (d) ion flux toward the wall as a function of gas pressure with four boundary condition assumptions: 1) $\gamma = 0$ (completely absorbing boundary), 2) $\gamma = 0.116$ (uncharged electrode surface), 3) $\gamma = 0.382$ (fully charged electrode surface) and 4) charge–controlled ISEY.

The ionization rate in the center of the bulk plasma and at the bulk–sheath interface is displayed in Fig. 10. It can be observed that there is a significant difference between the trend of ionization rate in the center of the discharge domain and at the bulk–sheath interface. In the bulk plasma center, the ionization rate rises at a low pressure (below 15 Pa) and declines after the gas pressure reaches a critical value. While at the sheath edge, the ionization rate continues to increase with the gas pressure. At low pressures, when the background gas becomes denser, there are more gas atoms for the ionization, increasing the ionization rate both in the bulk plasma center and at the bulk–sheath interface. In contrast, as the pressure keeps going up, the background gas becomes too dense, frequent collisions result in the decrease of electron energy in the bulk plasma center, thus decreasing the ionization rate there. At the bulk–sheath interface, the ionization is mainly induced by energetic electrons accelerated by electric field in the sheath. With the increase of the gas pressure, more energetic electrons are generated, leading to a higher ionization rate. In addition, at all gas pressures, the introduction of charge traps leads to a higher ionization rate not only in the bulk plasma center but also at the bulk–sheath interface due to the plasma generated by the SEs. Meanwhile, as presented in the Fig. 10, the curves of the ionization rate in the bulk plasma center and at the bulk–sheath interface intersect at about $p = 35$ Pa for an absorbing boundary, at $p = 30$ Pa for the charge–controlled ISEY. Above this intersection, the rate at the bulk–sheath interface becomes larger than the one in the center of the discharge domain and plays a more and more important role as the gas pressure increases. This change presents the shift of the maximum ionization rate. The intersection point varies with the boundary conditions, indicating the influence of charge traps on the ionization distribution.

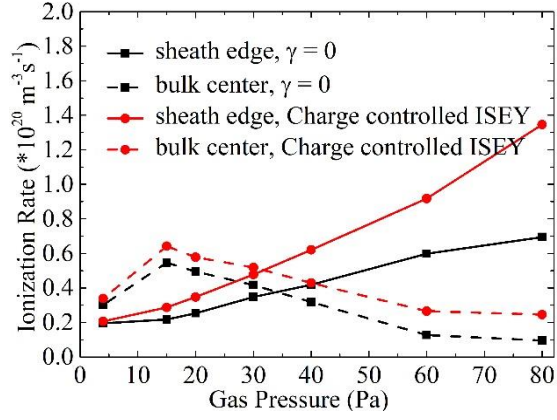

Fig. 10 Time–averaged ionization rate at the sheath edge and in the bulk plasma center as a function of gas pressure

**C. Discussion of asymmetric boundary condition**

In this section, we would like to investigate the impact of asymmetric boundary conditions on plasma discharge with the particle simulation code mentioned above. The $SiO_2$ electrode is employed only at grounded side and the powered electrode is set to a completely absorbing boundary. The thickness of $SiO_2$ and the distance between the two electrodes remain the same.

In a similar fashion to Fig. 6 and 7, Fig. 11 presents the time–averaged spatial distribution of the a) ion density $n_i$, b) electric potential, c) mean electron energy, d) heating rate of electrons $P_e$, e) ion power density heating rate of ions, and f) ionization rate at different pressures with asymmetric boundary conditions.

As explained before, charge traps result in the increase of the ion density due to the ionization induced by SEs. This result can also be observed in Fig. 11 (a) and (e): charge traps increases the ionization rate, thus the ionization rate and ion density peak appear in front of grounded electrode, where the charge–controlled ISEY is employed. The impact of gas pressure on ionization is also noticeable in Fig. 11. At a low pressure ($p$ = 4 Pa), the ionization mainly happens in the bulk plasma center and it's virtually symmetric. That's because the ion flux toward the boundary is relatively small, thus the number of SEs is also small and inconsequential. From 4 Pa to 80 Pa, it's obvious that the maximum ionization rate gradually shifts toward the bulk–sheath interface. As the gas pressure goes up, the electron mean free path decreases, thus the electrons accelerated by the plasma sheath lose their energy before reaching the center of discharge domain. At the same time, the number of SEs significantly augments as a result of a greater ion flux, therefore the impact of charge traps becomes greater, leading to a more remarkable plasma asymmetry, in agreement of the results in Fig. 6 (a), (e) and 7 (a), (e).

In addition, the electric potential near the grounded electrode gets smaller and the peak shifts toward the powered electrode. This phenomenon fits the case of symmetric boundary conditions in which the electric potential decreases with charge traps, as shown in Fig. 6 (b) and 7 (b). Meanwhile, it is in agreement with the theoretical analysis presented in Section III (Eq. (18) - (28)).

The time–averaged electron heating rate has similar peak at both sides, but the heating rate near boundaries are influenced by the asymmetric boundary conditions as presented in Fig. 11 (c). The electron heating rate near the grounded electrode is higher than the other side due to the implication of charge traps. For the ion case, the peak of time–averaged heating rate shifts toward the grounded side as shown in Fig. 11 (d). This asymmetry shows agreement with Fig. 6 (d) and 7 (d) in which the charge traps multiply the peak of ion heating rate near electrodes compared with the absorbing boundary. The reason could be the increasing ion flux due to the ion density change, the same as mentioned in Section B. The distribution of the mean electron energy is more related with the gas pressure rather than the ISEE, as shown in Fig. 11 (f).

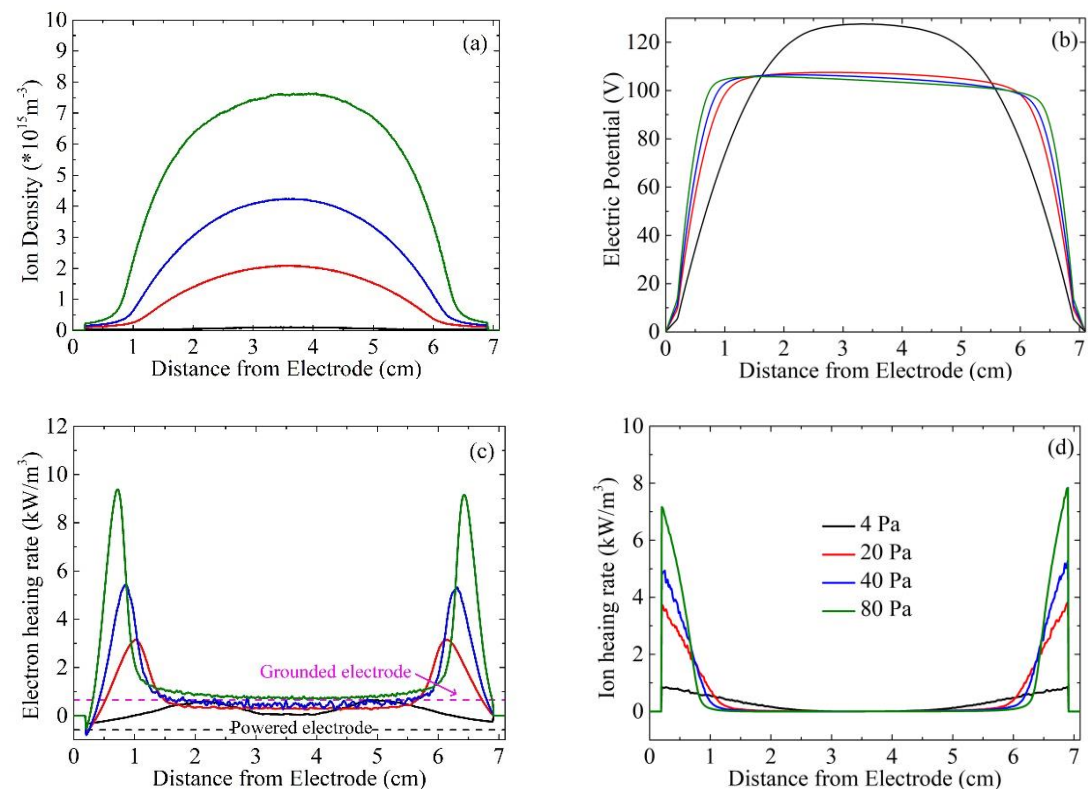

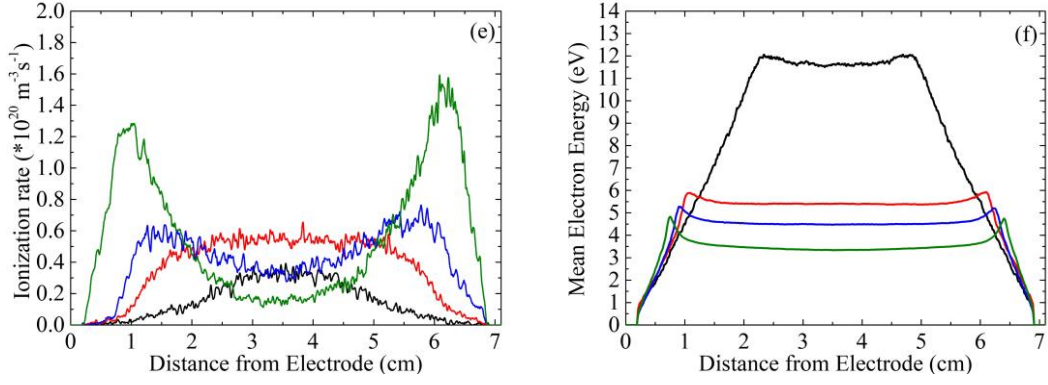

Fig. 11 Time–averaged spatial distribution of the a) ion density $n_i$, b) electric potential, c) mean electron energy, d) heating rate of electrons $P_e$, e) heating rate of ions $P_i$ and f) ionization rate with asymmetric boundary conditions at various pressures

**V. Conclusion**

In this work, the impact of charge traps on plasma properties is investigated theoretically and with a PIC/MCC simulation in Helium gas at various pressures, for 2mm $SiO_2$ electrodes. The discharges are driven at a frequency of 13.56 MHz and at a voltage amplitude of 210 V. Both symmetric and asymmetric boundary conditions are considered.

To reveal the impact of charge traps, a realistic ISEY model is employed, involving the impact of electron traps in the electrode material and surface charge accumulation. A realistic electron distribution in the valence band obtained from experiments is implemented. The resulting ISEY is a function of the surface charge accumulation. When the electrode is negatively charged, which is the normal case in low–pressure plasma discharges, the ISEY significantly increases compared with uncharged case. This result is expected as it is easier for electrons to escape when the material is negatively charged.

In addition, the charge traps are proved helpful in lowering the sheath potential, augmenting the power loss and stabilizing plasmas. When charge traps are considered, the sheath potential becomes smaller with the accumulation of the surface charge. Meanwhile, the analysis of the power balance indicates that charge traps increase the power loss by producing more electrons. Charge traps also stabilize the plasma, because $\frac{\partial \Gamma_e}{\partial \varphi_{sh}}$ decreases with the charge–controlled ISEY.

As discussed throughout and verified by the simulation results, the charge traps influence various plasma properties. They lead to a higher plasma density and ionization rate, a lower sheath potential, increase the power loss and affect the ionization rate. Furthermore, the impact of charge traps is related to gas pressure: plasma properties are less influenced by the charge traps in the electrode material at low pressures, while the effect of charge traps becomes greater as the gas pressure increases.

When considering a setup with asymmetric boundary conditions, the resulting asymmetry in plasma characteristics is not noticeable at low pressures but an obvious asymmetry appears at $p$ = 80 Pa. The simulation of asymmetry also provides an option for creating an asymmetric plasma with geometrically symmetric structures which is especially useful in $\gamma$–mode discharges.

## VI. Acknowledgments

This work was supported by the National Natural Science Foundation of China (Grant No. 51827809, 51777164); the Fundamental Research Funds for the Central Universities, China (Grant Nos. xtr042019009 and PY3A083).